\documentclass[12pt]{iopart}
\usepackage{harvard}
\citationmode{abbr}
\usepackage{graphicx}
\usepackage{subcaption}
\usepackage{hyperref}
\usepackage[utf8]{inputenc}
\usepackage[export]{adjustbox}
\usepackage{wrapfig}
\graphicspath{{./}{figures/}}
\begin{document}

\title[Probing the atmospheric precipitable water vapor with SOFIA, Part I]{Probing the atmospheric precipitable water vapor with SOFIA, Part I, Measurements of the Water vapor overburden with FIFI-LS}

\author{C Fischer$^1$, C Iserlohe$^1$, W Vacca$^2$ , D Fadda$^2$, S Colditz$^1$, N Fischer$^1$ and A Krabbe$^1$}\address{$^1$ Deutsches SOFIA Institut, University of Stuttgart,Pfaffenwaldring 29, D-70569 Stuttgart, Germany}\address{$^2$ USRA SOFIA, NASA Ames Research Center, MS N232-12, Moffett Field, CA 94035-1000, USA}\ead{fischer@dsi.uni-stuttgart.de}

\vspace{10pt}
\begin{indented}
\item[]January 2021
\end{indented}

\begin{abstract}
We report on the measurements of telluric water vapor made with the instrument FIFI-LS on SOFIA. Since November 2018, FIFI-LS has measured the water vapor overburden with the same measurement setup on each science flight with about 10 data points throughout the flight. This created a large sample of 469 measurements at different locations, flight altitudes and seasons. The paper describes the measurement principle in detail and provides some trend analysis on the 3 parameters. This presents the first systematic analysis with SOFIA based on in situ observations.
\end{abstract}

%
%
%
%
%

\section{Introduction} \label{sec:intro}

Airborne astronomical observations are carried out through a residual atmosphere with a transmission between 0 and $\sim$98$\%$ depending on the wavelength and local atmospheric conditions. Particularly for spectroscopic observations at certain wavelengths, knowledge of the current atmospheric transmission is essential to achieve good calibration. For example, for the important [OI] fine structure line at $\sim$63 $\mu$m the uncertainty in the correction of the atmospheric absorption along the line of sight can easily reach a factor of 2 \cite{Erickson98}. The water vapor overburden at a certain time, location and altitude varies with the weather.

Here we report on measurements of the precipitable water vapour (PWV) carried out with the Field-Imaging Far-Infrared Line Spectrometer (FIFI-LS) \cite{Fischer18} \cite{Colditz18} on board the Stratospheric Observatory for Infrared Astronomy (SOFIA) \cite{Erickson93}. FIFI-LS is an integral-field imaging spectrometer that provides simultaneous observations in two channels: the blue channel covering 51-125 $\mu$m and the red channel covering 115-203 $\mu$m.  Each channel consists of an overlapping 5$\times$5 pixel footprint on the sky, where the pixel size is $6"$ and $12"$ for the blue and red channels, respectively.  We refer to each of these spatial pixels as ``spaxels''. The optics within FIFI-LS rearranges the 25 spaxels into a pseudo-slit, and the light impinging on each spaxel is then dispersed spectrally (using a grating) over 16 pixels. This generates an integral-field data cube for each observation.  The spectral resolution is wavelength dependent ranging from $R = \lambda$/$\Delta\lambda$ = 500 to 2000.

To determine the water vapor overburden the instrument configuration is set to measure multiple telluric water lines in emission without any background subtraction. Based on models of the atmospheric transmission (e.g. \citeasnoun{Lord92}), it is known that the characteristic shape and strength of these lines depends strongly on the water vapor overburden. A model of the emission spectrum is fitted to the observed spectrum to determine the water vapor overburden without the need to calibrate the data. Both channels of the instrument are configured to observe telluric water lines simultaneously to provide two independent measurements. About 10 measurements are carried out during each flight, usually while the telescope is set up for the next target. These measurements provides water vapor estimates for the telluric correction of science data as well as to inform strategic decisions regarding data acquisition in flight. These observations represent a substantial improvement over the currently available water vapor information both in flight as well as post-flight.  For SOFIA, predictions of the water vapor overburden at the aircraft's location and altitude during a flight are available from a model implemented in the flight planning software \cite{Leppik18} but this provides only a statistical estimate without any measured parameters from the atmosphere to create a reliable forecast.

\begin{figure}[h]

\includegraphics[width=1\linewidth]{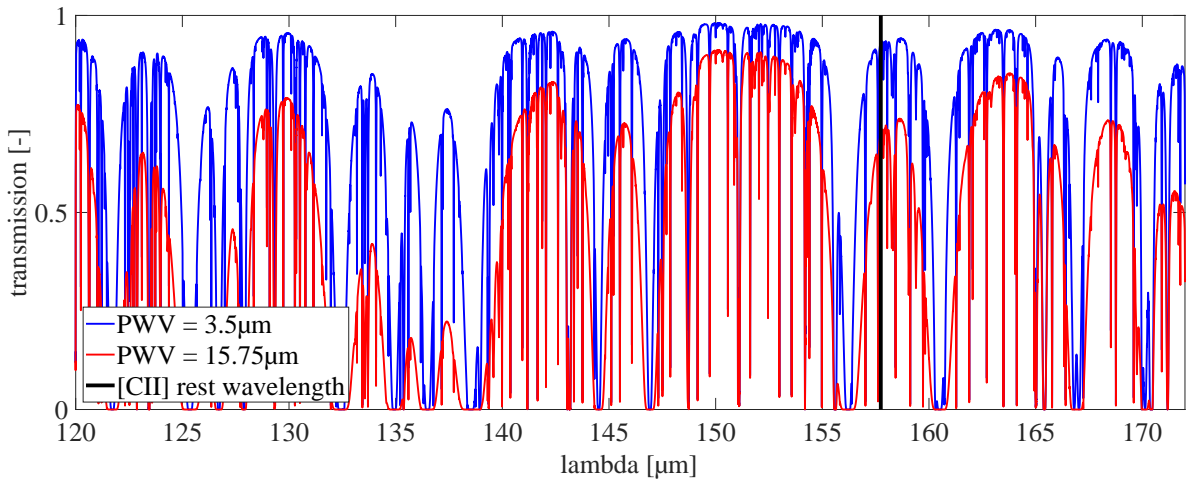}

\caption{Atmospheric transmission between 120 $\mu$m and 172 $\mu$m modeled with ATRAN for a pressure altitude of 39000 ft and a telescope elevation of 40$^{\circ}$. The 2 curves show the transmission for the highest (15.75 $\mu$m) and lowest (3.5 $\mu$m) PWV values measured by FIFI-LS at this altitude. The black line marks the rest wavelength of [CII].}
\label{fig:atm}
\end{figure}

During the reduction of FIFI-LS science data, a typical value\footnote{https://dcs.arc.nasa.gov/proposalDevelopment/SITE/index.js} of the water vapor overburden for the pressure altitude\footnote{When referring to altitude throughout the paper, we always refer to pressure altitude, since SOFIA flies on levels of constant pressure altitude.} at which the observations were obtained is used as an input parameter into an atmospheric model to generate the transmission curve used to correct the spectra for atmospheric absorption. This procedure can be problematic for observations in spectral regions where the telluric transmission strongly depends on the level of water vapor, like at the location of the [OI] line mentioned above. Also, for extra-galactic sources the line can easily get redshifted into a problematic region of the spectrum. One of the spectral lines observed most frequently with FIFI-LS is the [CII] line at a rest wavelength of 157.741 $\mu$m. The line has been detected with FIFI-LS up to a redshift of $\sim$0.085 or $\sim$171.15 $\mu$m \cite{Pusching20}. Within this wavelength range there are multiple absorption features in the atmosphere with a strong dependence of the transmission on the water vapor overburden. The atmospheric transmission as modeled with ATRAN (see section \ref{sec:principle}), is shown in figure \ref{fig:atm}. The altitude was set to 39000 ft (representing the lowest typical observing altitude for SOFIA) and the telescope elevation of 40$^{\circ}$, the center of the 22$^{\circ}$ to 58$^{\circ}$ range. The transmission curves correspond to the highest (15.75 $\mu$m) and the lowest (3.5 $\mu$m) (see section \ref{fig:results}) PWV values ever measured in our data sample at 39000 ft. The relative offset between the curves clearly demonstrates the challenge for a quality atmospheric calibration. The rest wavelength of the [CII] line is also marked. Within the range of redshifts observed with SOFIA/FIFI-LS there are numerous deep absorption features with a strong dependency on the PWV value (e.g., at $\sim$161 $\mu$m or $\sim$167 $\mu$m). If the redshifted line is located on the edge of such a feature the transmission can easily drop by a factor of 2 or more. This illustrates the need to accurately determine the water vapor overburden to achieve good calibration of the science data across a wide range of wavelengths.

In addition to the calibration benefits, knowledge of the seasonal or local trends of water vapor also provides an opportunity to increase the observing efficiency of the observatory. Because integration times for astronomical observations are calculated using an estimate of the water vapor overburden, a better knowledge of the water vapor and potential trends may be used to refine integration times based on the flight plan parameters (season, location, altitude, elevation). Also, observations at wavelengths more sensitive to water vapor could be scheduled  such as to to minimize the risk of high water vapor.

Water vapor has been measured with SOFIAs predecessor,  the Kuiper airborne observatory (KAO), in a variety of studies. In \citeasnoun{Nolt79} and \citeasnoun{nolt80} water vapor was measured on 39 flights at 41000 ft and some additional data points at an altitude of 45000 ft in summer and fall on flights out of the NASA Ames Research Center in northern California and also out of Hawaii during summer. They derived a median zenith value at 41000 ft of 6.5 $\mu$m for California and 8.5 $\mu$m for Hawaii with values ranging from 4-11 $\mu$m. For 45000 feet, the Median was 4.5 $\mu$m out of Hawaii with values ranging from 3-8 $\mu$m. While trends with location and with altitude were clearly demonstrated, the number of data points was very limited. Similar values were reported in \citeasnoun{erickson79} with overburdens between 5 and 30 $\mu$m, measured on 5 flights with strong variation in between flights. PWV values in the same range, 7 to 15 $\mu$m at 41000 ft, were reported by \citeasnoun{Kuhn82}. Very low values of water vapor were reported by \citeasnoun{lord96L} with values below 3 $\mu$m at 43000 ft and below 2 $\mu$m at 45000 ft. From all these observations it is clear that the overburden varies not only with position, season and altitude but also with ever changing weather on a flight to flight basis.

To better capture the influence of position and season on the PWV, a different approach was taken by \citeasnoun{1Haas98}. Satellite data were used to create maps of water vapor overburden at 41000 ft for the four meteorological seasons. Their maps show a clear distinction between the seasons with low values (below 10 $\mu$m) in the winter (December - February) for the area accessible in a science flight from Palmdale (see figure \ref{fig:results} for the area covered by the FIFI-LS flights). The range of values found were close to those determined from the KAO measurements with values between 4 and 24 $\mu$m. They are very similar during the spring (March - May) with values slightly higher but still well below 10 $\mu$m in the relevant area for flights out of Palmdale. Water vapor increases substantially in the summer (June - August) especially over the southern and southeastern USA. PWV values measured in fall are between those estimated for the summer and spring. In addition, during all seasons there is a north/south trend present. The PWV values increase towards the south with a strong upwards trend in the tropics south of ~20$^{\circ}$N. They does not provide any data on the expected flight to flight variation. Since they found a strong dependence of the PWV values  on the meteorological seasons we will also use these for own trend analysis. We will also check for the generally observed trend with latitude (lower water vapor overburden towards the north). In \citeasnoun{Iserlohe21:inpress-b} a similar approach, also using satellite data, is pursued for the the FIFI-LS flights to make PWV values available for any point on the flight path, not only when a measurement is performed roughly once per hour.

\section{Data acquisition and reduction} \label{sec:reduction}
The data were taken as part of regular FIFI-LS observing flights with SOFIA between November 2018 and September 2020 on 39 flights during six flight series. Data were taken during all four meteorological seasons, with six flights in winter, 12 flights in spring, three flights in the summer and 18 flights in the fall. Details about the dates are given in table \ref{tab:flights}. In total 469 measurements were obtained. Data were taken during the telescope setup leg at the beginning of each flight and during the setup on target at the beginning of each leg. (A leg is the observation of a single source or region in the sky. A typical science leg during a 10h flight is between 30 min and 3 h long.)   Science observations were paused to take PWV measurements after the aircraft climbed in altitude or when the quick look reduction of the data indicated changing background levels in the raw files. 

\begin{table}
\caption{\label{tab:flights}Details about the flights}
\begin{indented}

\item[]\begin{tabular}{@{}lll}


\br
flight series&date range&number of flights\\
\mr
OC6M & 06.11.2018 - 09.11.2018  & 4\\
OC6U & 27.02.2019 - 02.03.2019 & 3 \\
OC7A & 01.05.2019 - 17.05.2019 & 11 \\
OC7H & 30.10.2019 - 14.11.2019 & 10 \\
OC7L & 25.02.2020 - 28.02.2020 & 4  \\
OC8B & 17.08.2020 - 04.09.2020 & 7 \\
\br
\end{tabular}
\end{indented}
\end{table}

During a science flight SOFIA starts observing at an  altitude of 38000 ft and then climbs throughout the flight in increments of 1000 ft as the aircraft becomes lighter. The climbs are usually planned based on the assumed take off weight, the fuel burn rate and the standard atmospheric temperature. Deviations are then caused by varying local temperatures, air traffic control requirements and turbulence. On most flights the maximum  altitude is 43000 ft, while on some flights depending on fuel load and weather conditions up to 45000 ft can be reached. Since only very few measurement points are available at altitudes above 43000 ft and water vapor values are generally very low, we combine values obtained at 44000f and 45000 ft in the data analysis for better statistics.  

\begin{figure}[h]

\includegraphics[width=1\linewidth]{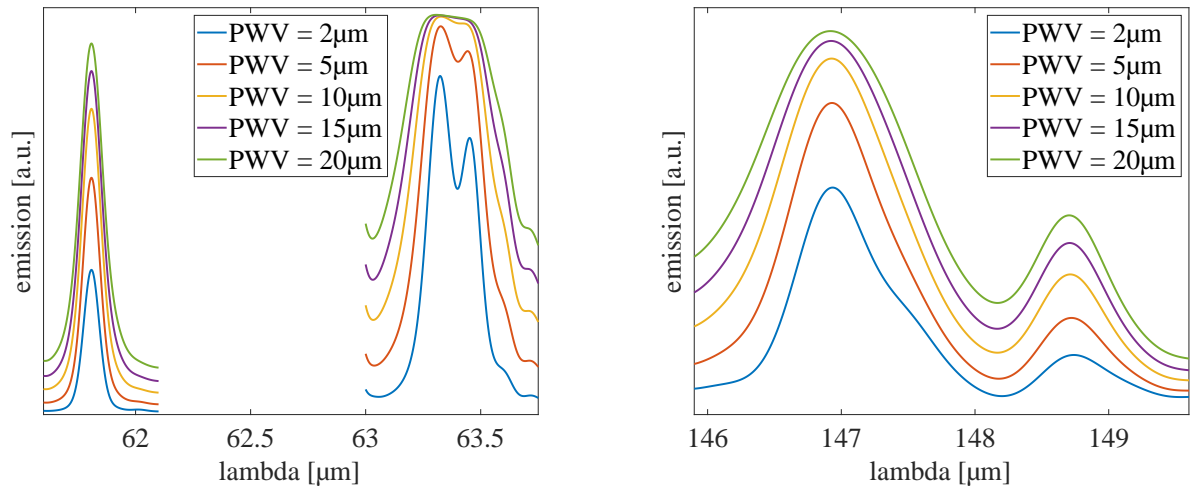}

\caption{Modelled atmospheric emission spectra  based on equation \ref{eq:1} for different values of water vapor overburden for both FIFI-LS channels in the wavelength range used for the PWV measurement. Blue Channel model spectra are shown on the left panel, red channel model spectra on the right panel. All spectra shown are at for 41000 ft altitude, 40$^{\circ}$ telescope elevation and 200K temperature. The spectra are convolved with the instruments spectral resolution in the wavelength range.}
\label{fig:spectra}
\end{figure}

The data are taken without chopping or nodding and a total of five integrations centered at different wavelengths are required to provide the spectral range shown in figure \ref{fig:spectra}. The integration time at each spectral setting was 3 s yielding $\sim$1 min to complete a full measurement including all overheads. Data are generally taken with the dichroic filter (\cite{Fischer18}) used in the current or upcoming science observation to avoid excessive use of the filter changer mechanism.  The data are reduced with the FIFI-LS data reduction pipeline \cite{vacca20}, with chop and nod subtractions disabled. Then all the spatial pixels are collapsed to provide a single spectrum to be analyzed.

\section{Measurement principle} \label{sec:principle}
We have chosen parts of the spectrum within the range of FIFI-LS that contain telluric water lines which are sensitive in their relative amplitudes and widths to changes in water vapor within the range typically encountered by SOFIA. We chose features that require limited spectral bandwidth to minimize the time required to obtain the measurements in order not to significantly impact the observing efficiency. We use ATRAN \cite{Lord92} to model the transmission of the atmosphere. The model calculates the transmission of the atmosphere as a function of wavelength, pressure altitude, PWV in the zenith\footnote{In order to keep PWV values comparable throughout the paper, we always refer to PWV in the zenith, not along the line of sight of the telescope.} and telescope elevation. We use the transmission curves from the ATRAN model and transform them into an emission model (E) using equation \ref{eq:1}. The transmission from the ATRAN model is combined with a black body to model sky emission at the wavelength range within the spectral region. Since the measured data is not calibrated all multiplicative constants have been removed. For the temperature $T$ the static air temperature measured by the aircraft is used, along with the pressure altitude (alt) and telescope elevation (el) at the time of observation.
\begin{equation} \label{eq:1}
E (\lambda,T,PWV_{zenith},alt,el) =  \frac{1-Tr_{ATRAN}(\lambda,PWV_{zenith},alt,el)}{\lambda^5e^{\frac{hc}{\lambda kT}}-1}
\end{equation}
Here $Tr_{ATRAN}(\lambda,PWV_{zenith},alt,el)$ is the atmospheric transmission from ATRAN. The emission based on equation \ref{eq:1} in the selected spectral region is shown in figure \ref{fig:spectra} for multiple distinct values of water vapor. Since the atmosphere is observed without any background subtraction, the emission model is fitted to the measured signal using equation \ref{eq:1} augmented by an offset and a first order polynomial to account for offsets and scaling (equation \ref{eq:2}).   
\begin{equation} \label{eq:2}
EmissionModelFit(\lambda) =  a + b*\lambda + c*E(\lambda,T,PWV_{zenith},alt,el)
\end{equation}
The values a, b and c are then fitted to match the measured data. The sensitivity of the chosen emission lines to varying PWV values is shown in figure \ref{fig:spectra}. In the blue channel the water line at $\sim$61.8 $\mu$m grows relative to the doublet line at $\sim$63.4 $\mu$m with increasing water vapor values. Also the flux minimum between the lines of this doublet becomes less pronounced for higher PWV values and the overall width increases of the doublet increases. A similar behavior is found in the red channel. The line at $\sim$148.7 $\mu$m grows relative to the one at $\sim$146.9 $\mu$m, which becomes wider for higher values of PWV.

\begin{figure}[h]

\begin{subfigure}{0.48\textwidth}
\includegraphics[width=1\linewidth]{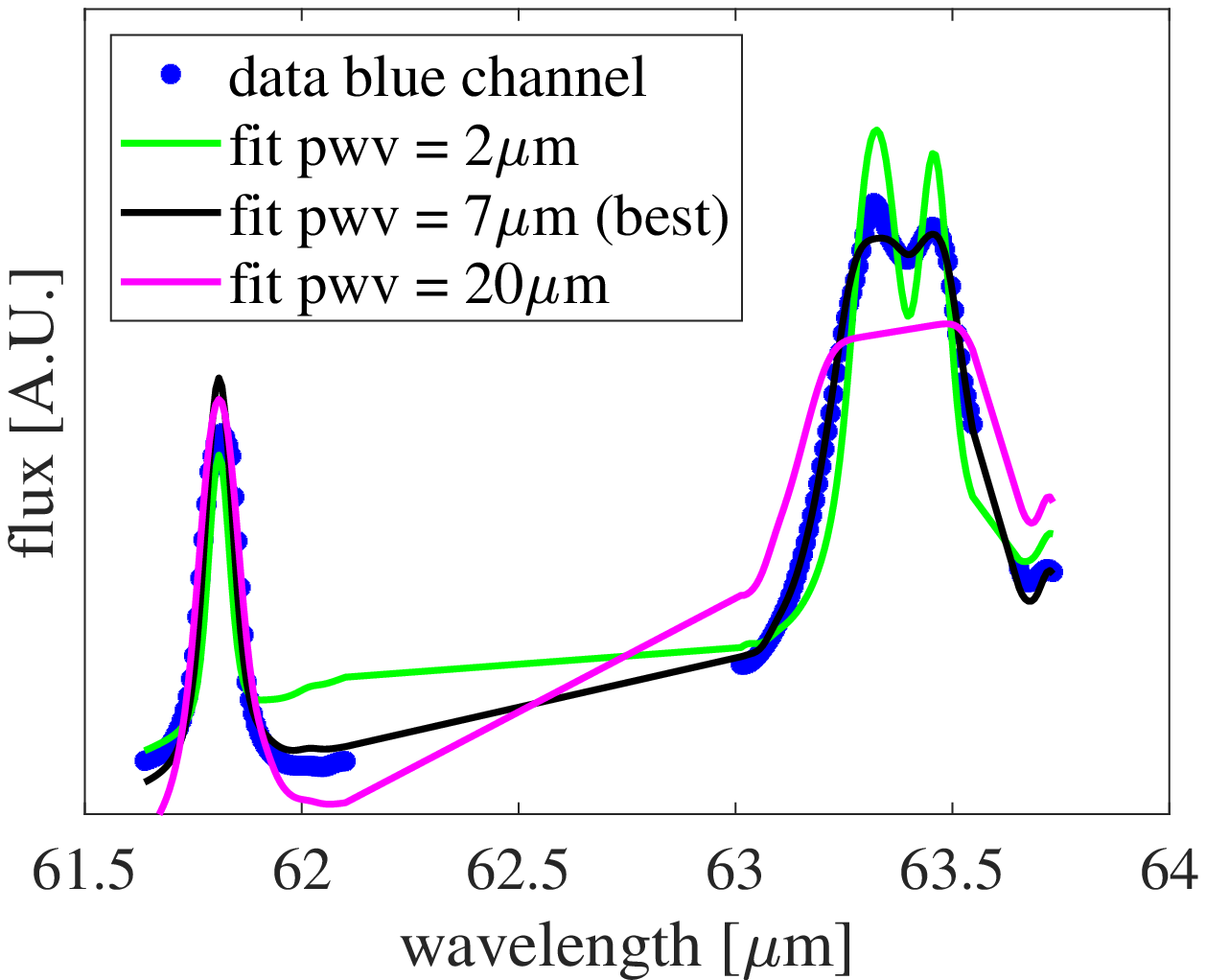}
\caption{blue channel}
\end{subfigure}
\begin{subfigure}{0.48\textwidth}
\includegraphics[width=1\linewidth]{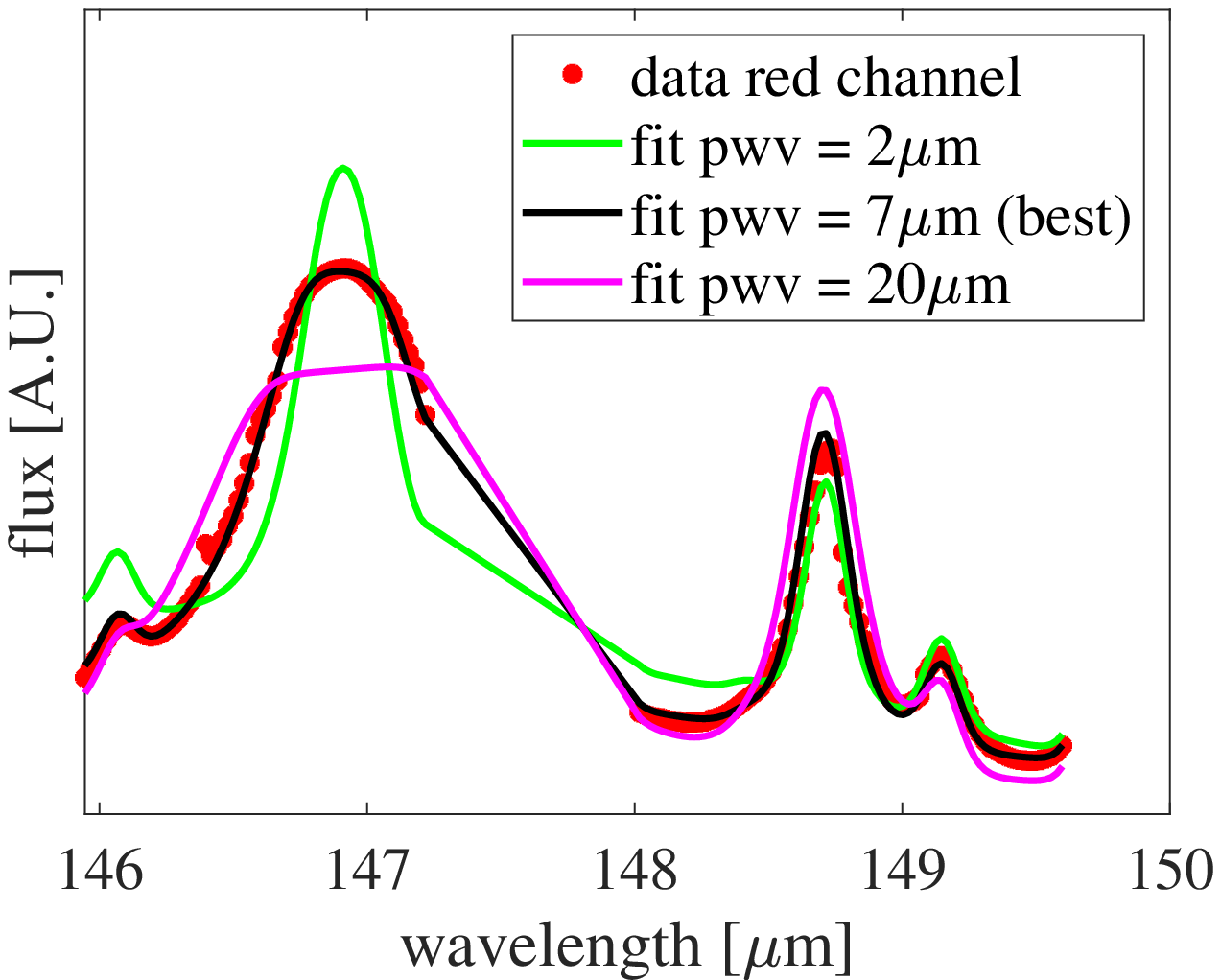}
\caption{red channel}
\end{subfigure}

\begin{indented}
\item[]\begin{subfigure}{0.48\textwidth}
\includegraphics[width=1\linewidth]{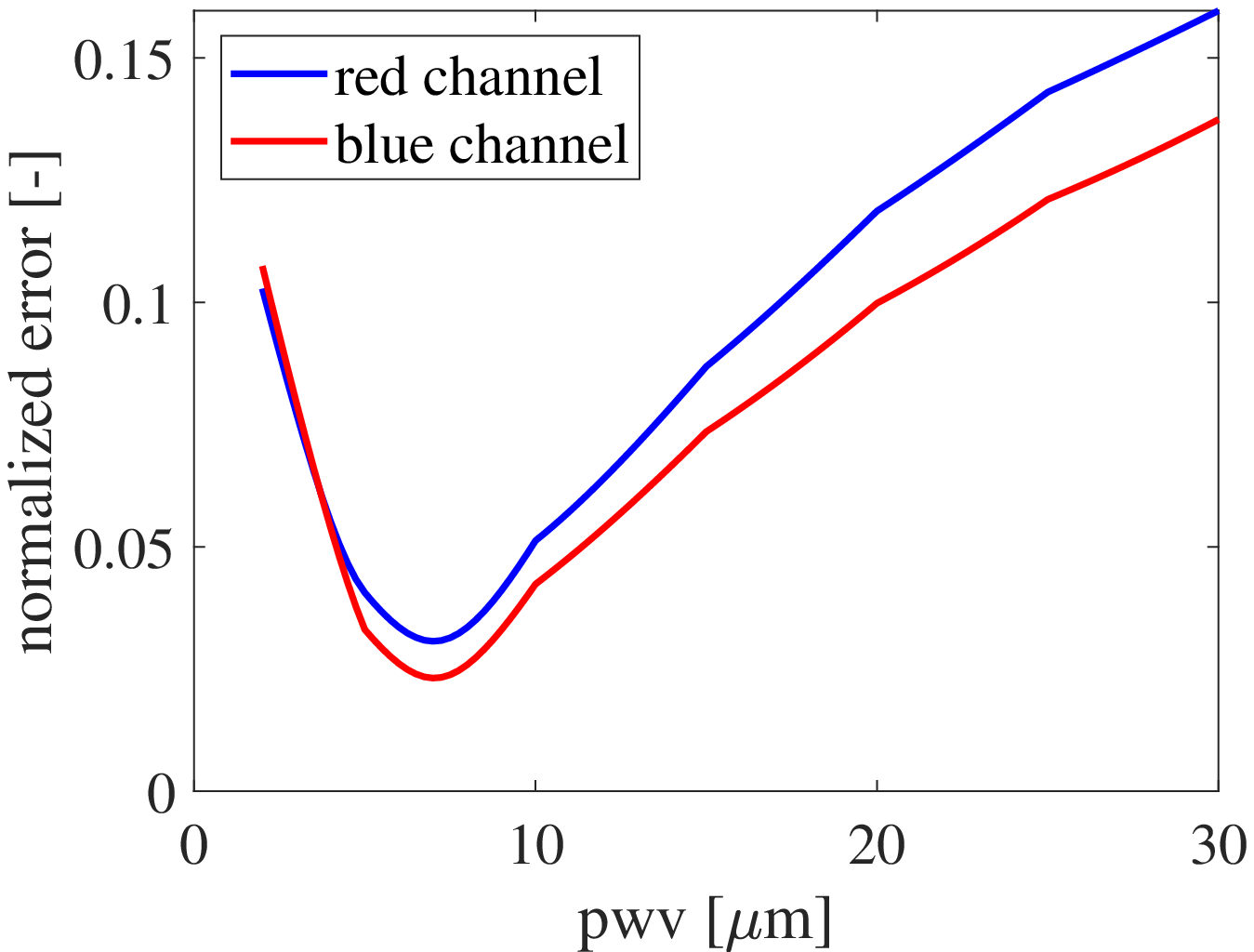} 
\caption{normalized error}
\end{subfigure}
\end{indented}

\caption{Example of a emission model fit: (a) The measured data (blue) in the blue channel overlayed with the best fit (black) as well the fits for 2 $\mu$m (green) and 20 $\mu$m (pink) (b) The same for the red channel. (c) Normalized errors for both fits in the red and the blue channel show a minimum at 7 $\mu$m PWV. The normalized error is the standard deviation between the measured and the fitted spectrum normalized with the mean flux in the measured spectrum. The well defined coincidence of both minima at the same PWV level demonstrate the quality of the procedure.}
\label{fig:fit}
\end{figure}

\section{PWV determination} \label{sec:data}
ATRAN transmission curves are generated for this analysis over a large range of flight altitudes, zenith angles and water vapor overburdens. With the altitude and zenith angle for each observation, emission curves (equation \ref{eq:1}) are then generated for water vapor values between 2 and 30 $\mu$m in 0.25 $\mu$m increments. For each of the emission curves an emission model (equation \ref{eq:2}) is then fitted to the data. The standard deviation between the fitted and the measured signal is then normalized with the mean flux measured. The PWV value resulting in the minimum normalized error for the fit is the result of the measurement. Spectra taken in the blue and the red channel are processed independently resulting in 2 water vapor values that are then averaged for the final estimate. For 5 flights during the OC8B flight series the FIFI-LS blue channel was not operational, and therefore only the red channel value was used. An example of the fit results for SOFIA flight 563 from May 2019 (OC7A) flying at 41000 ft with the telescope at a zenith angle of 50$^{\circ}$ are shown in figure \ref{fig:fit}. For both channels the normalized error of the fit has a clear minimum at 7 $\mu$m water vapor as can be seen in figure \ref{fig:fit} (c). Given that the measurements for both channels are completely independent this shows how well our approach works to determine the PWV values. This is also shown by spectra shown in (a) for the blue channel and (b) for the red channel. For the lowest error PWV value of 7 $\mu$m it is possible to fit the measured signal well with equation \ref{eq:2}. In addition to the best fit at 7 $\mu$m water vapor the fits for 2 $\mu$m and 20 $\mu$m are shown for both channels in green. For those values the fit clearly does not match the measured spectra in either width or relative amplitude of the lines. Averaged over all 419 available data points with data from both channels the standard deviation between the red and the blue channel yielded 0.3 $\mu$m of water vapor or 7$\%$ in mean relative offset.

\begin{figure}[h]

\begin{subfigure}{0.52\textwidth}
\includegraphics[width=1\linewidth]{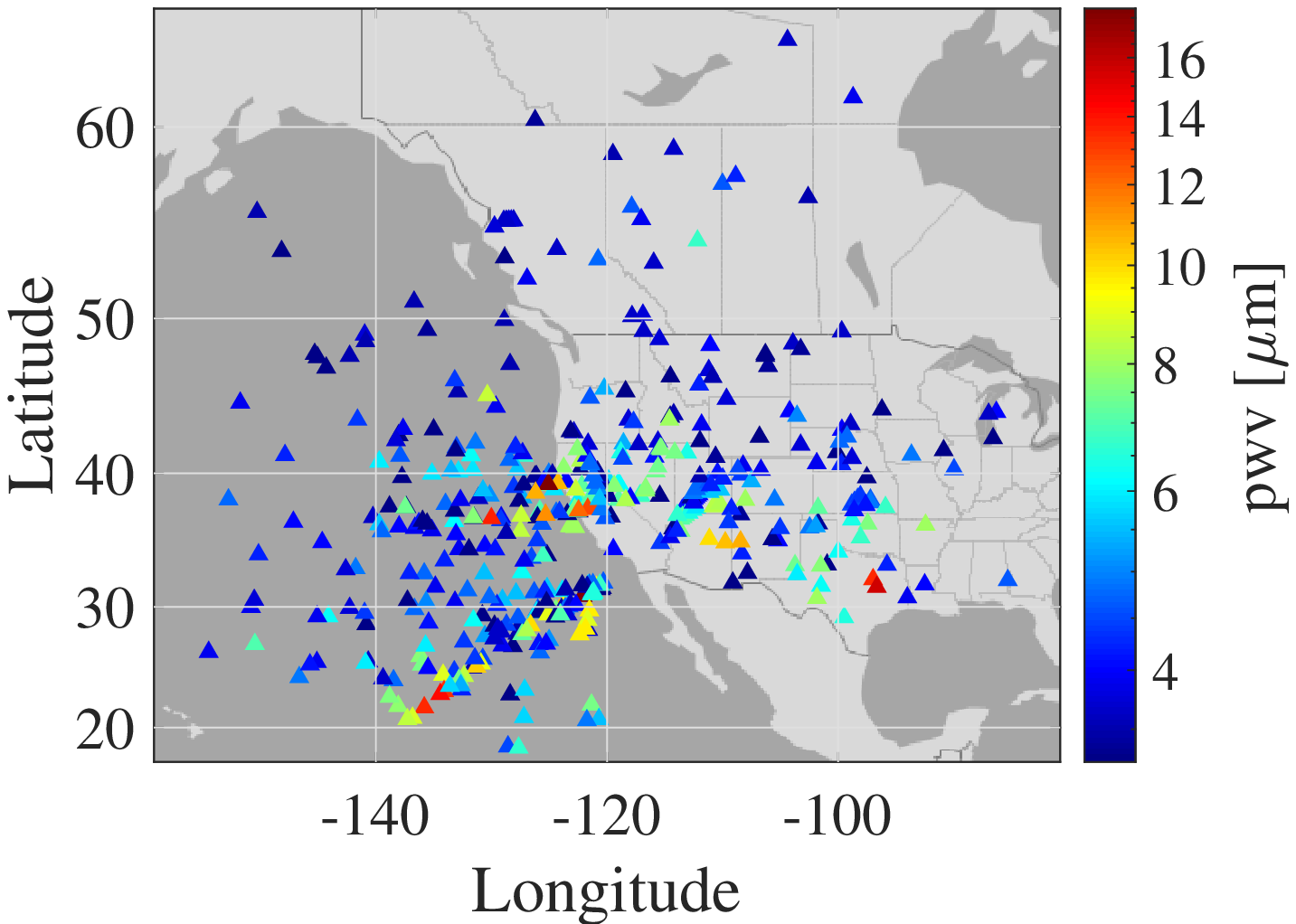}
\end{subfigure}
\begin{subfigure}{0.47\textwidth}
\includegraphics[width=1\linewidth]{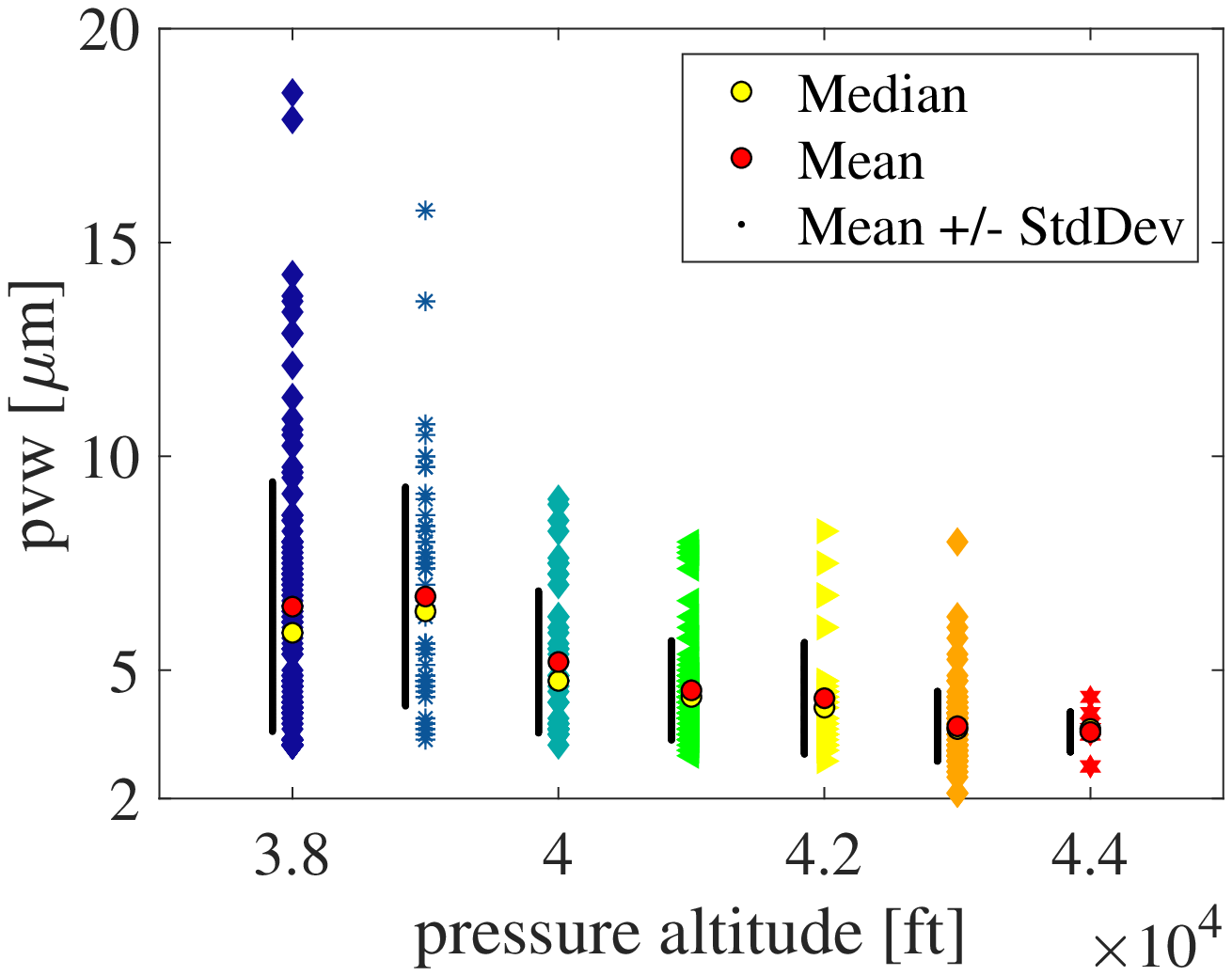}
\end{subfigure}

\caption{Results from the the water vapor measurements: (a) All measured values plotted over a map at the position they have been taken. (b) The same results separated by the pressure altitude they have been taken at. The yellow and red dots mark the median and mean values and the black bars indicate the standard deviation.}
\label{fig:results}
\end{figure}

\section{Results} \label{sec:results}

The results from all measurements are shown in figure \ref{fig:results}. The spatial map was generated with the geoplot function of matlab\footnote{\url{https://www.mathworks.com/help/matlab/ref/geoplot.html}}. The lowest measured value is 2 $\mu$m and the highest is 18.5 $\mu$m. Measurements are distributed over a wide area across the United States, Canada and the Pacific Ocean, even though they are clustered around southern California due to the fact that all flights start and end in Palmdale near Los Angeles ($\sim$35$^{\circ}$ S, $\sim$118$^{\circ}$ W). SOFIA does not perform observing flights over Mexico.

In figure \ref{fig:results} (b) the measurements are shown discriminated by altitude. The mean and median values for each altitude is also shown as well as the respectrive standard deviation. At higher altitudes at and above 41000 ft mean and median values are basically identical, while the median is $\sim$0.5-1 $\mu$m lower at 40000 ft and below. This shows that at the lower altitudes high PWV outliers become more relevant since they do increase the mean value compared to the median. A clear trend of water vapor decreasing with altitude is shown; the same trend is also seen in the standard deviations. Median values at 38000 ft and 39000 ft are only about 1 $\mu$m higher than at 41000 ft. However the spread of the values is much higher at lower altitude with $\sim$5.5 $\mu$m at 38000 ft and 39000 ft compared to 2 $\mu$m at 41000 ft. Water vapor is usually considered problematic for zenith values above 10 $\mu$m and good below 7 $\mu$m although such a characterization strongly depends on the wavelength observed as well a the zenith angle of the source. From figure \ref{fig:results} it is clear that for the majority (80.2$\%$) of the observations PWV is $< 7 \mu$m. Out of 469 values only 20 are at 10 $\mu$m or above and 14 of those were taken at 38000 ft, very early in the flights during the telescope setup leg. On most flights science observations do not begin until an altitude of 39000 ft is reached. Less than 2$\%$ of the data points (6 of 336 at 39000 ft and above) indicate problematic water vapor values.  Basically all PWV values at 40000 ft or above are well below 10 $\mu$m. The issue at the lower altitudes is not the typical PWV value but the higher spread of values as represented by the standard deviation reaching up close to about 10 $\mu$m with a significant number of high outliers above that. The lower flight levels are best characterized with generally good conditions with a larger risk of high water vapor conditions. 

\begin{figure}[h]

\begin{subfigure}{0.32\textwidth}
\includegraphics[width=1\linewidth]{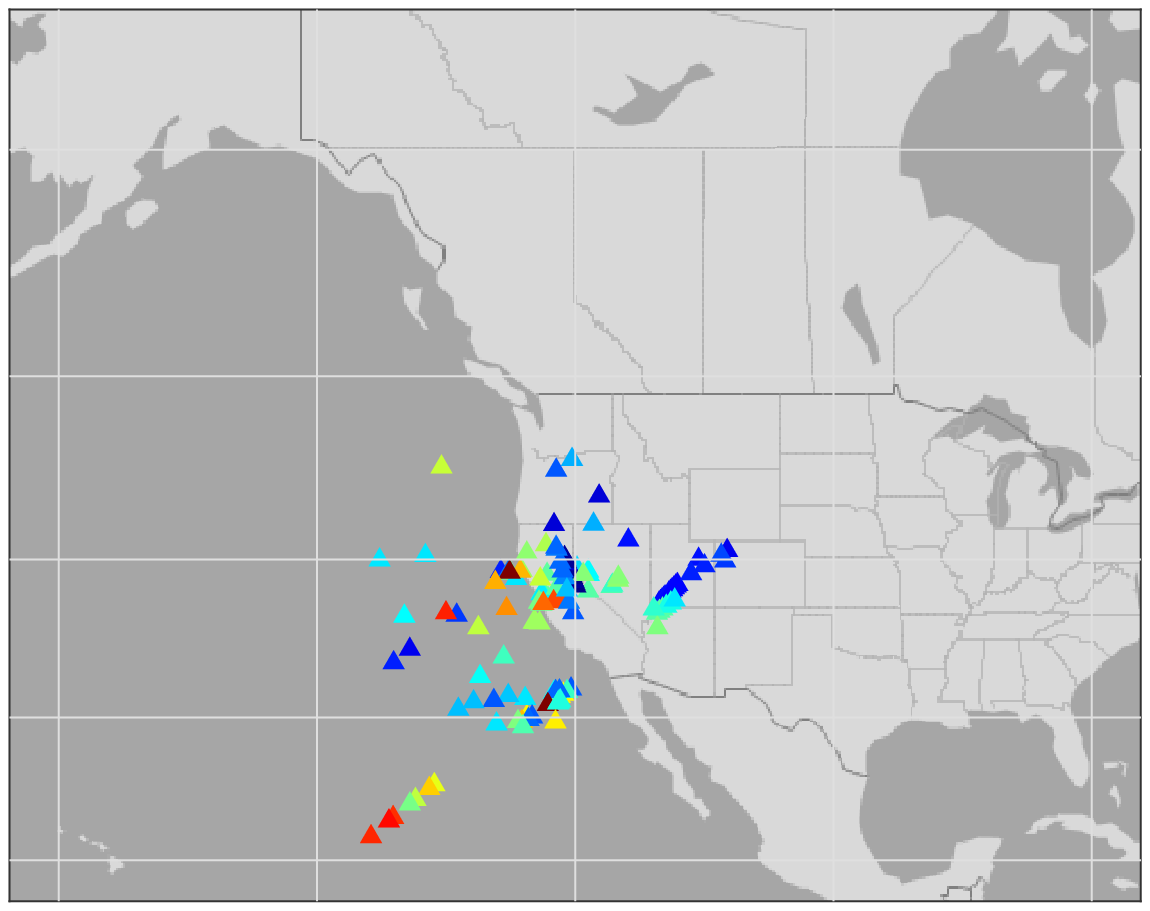}
\caption{38000 ft}
\end{subfigure}
\begin{subfigure}{0.32\textwidth}
\includegraphics[width=1\linewidth]{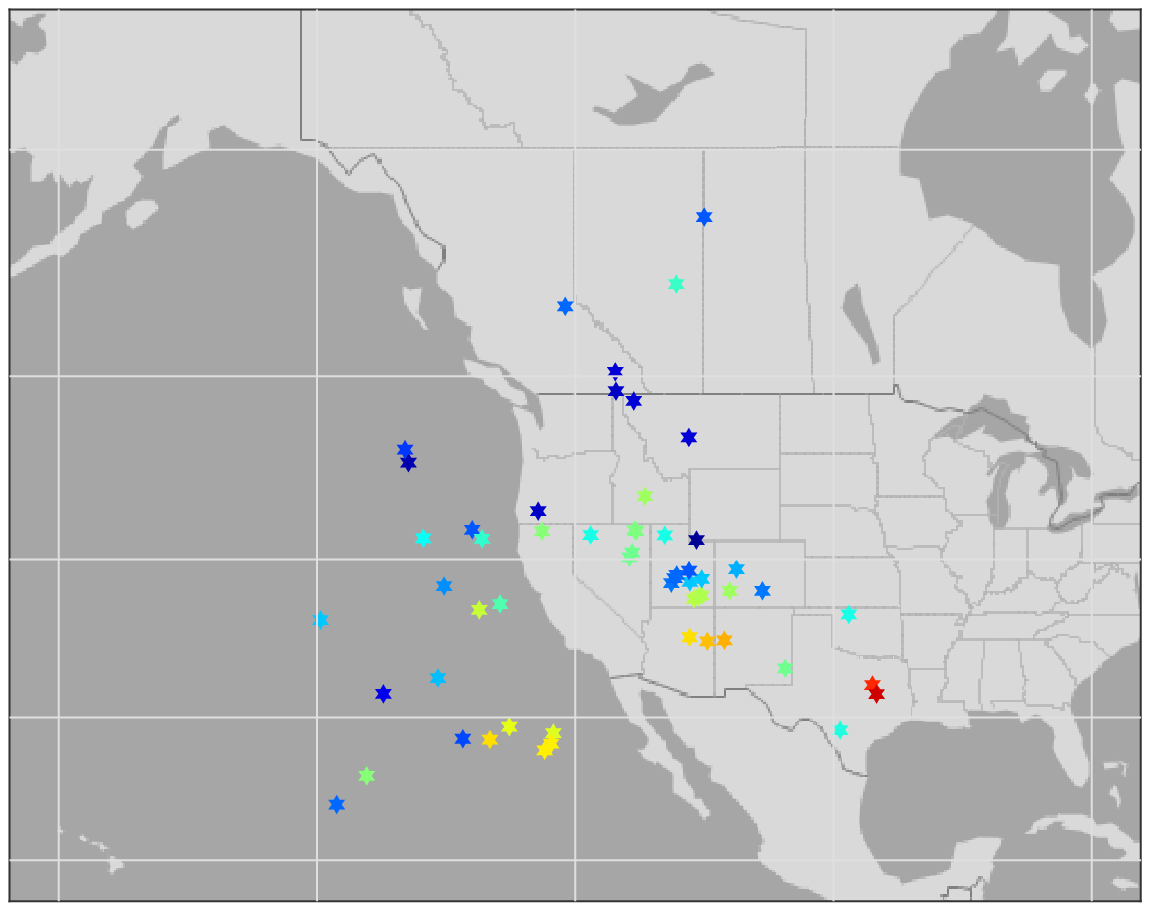}
\caption{39000 ft}
\end{subfigure}
\begin{subfigure}{0.32\textwidth}
\includegraphics[width=1\linewidth]{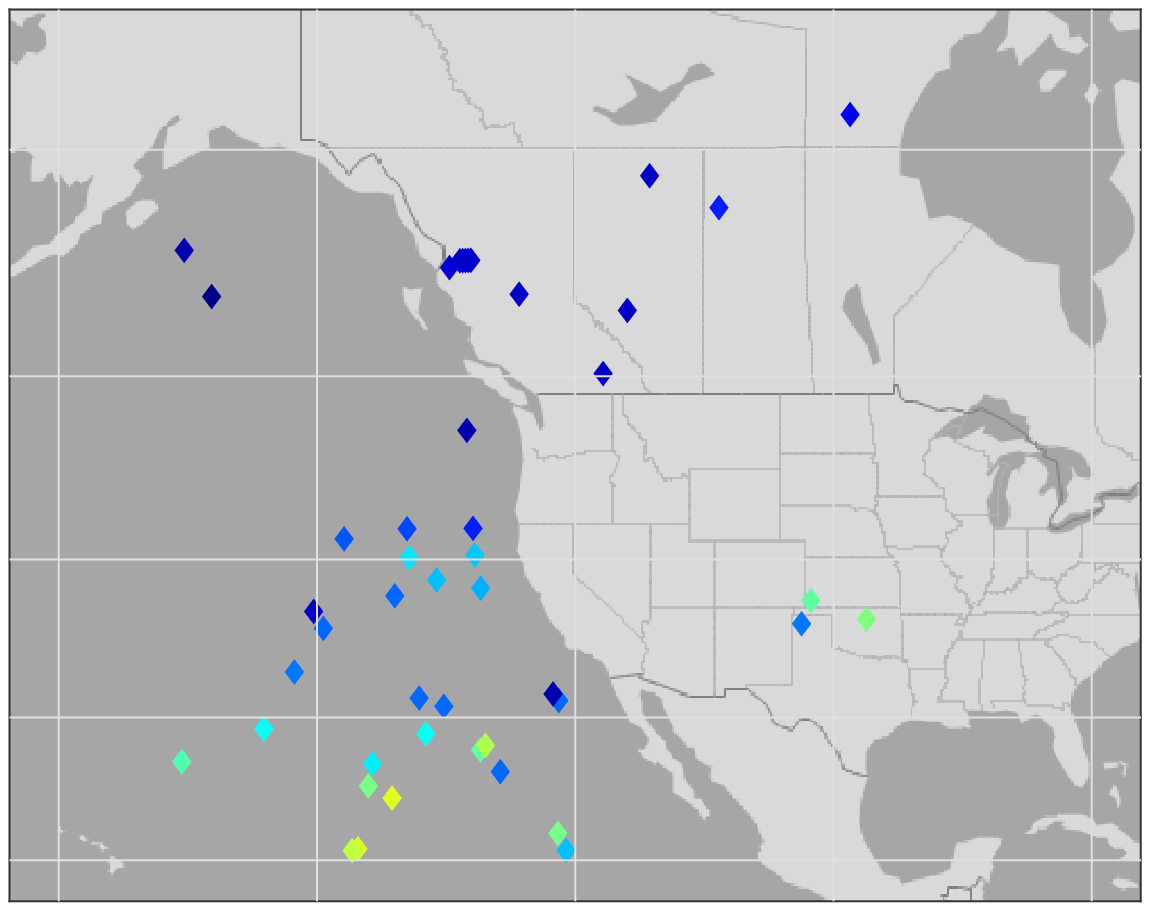}
\caption{40000 ft}
\end{subfigure}

\begin{subfigure}{0.32\textwidth}
\includegraphics[width=1\linewidth]{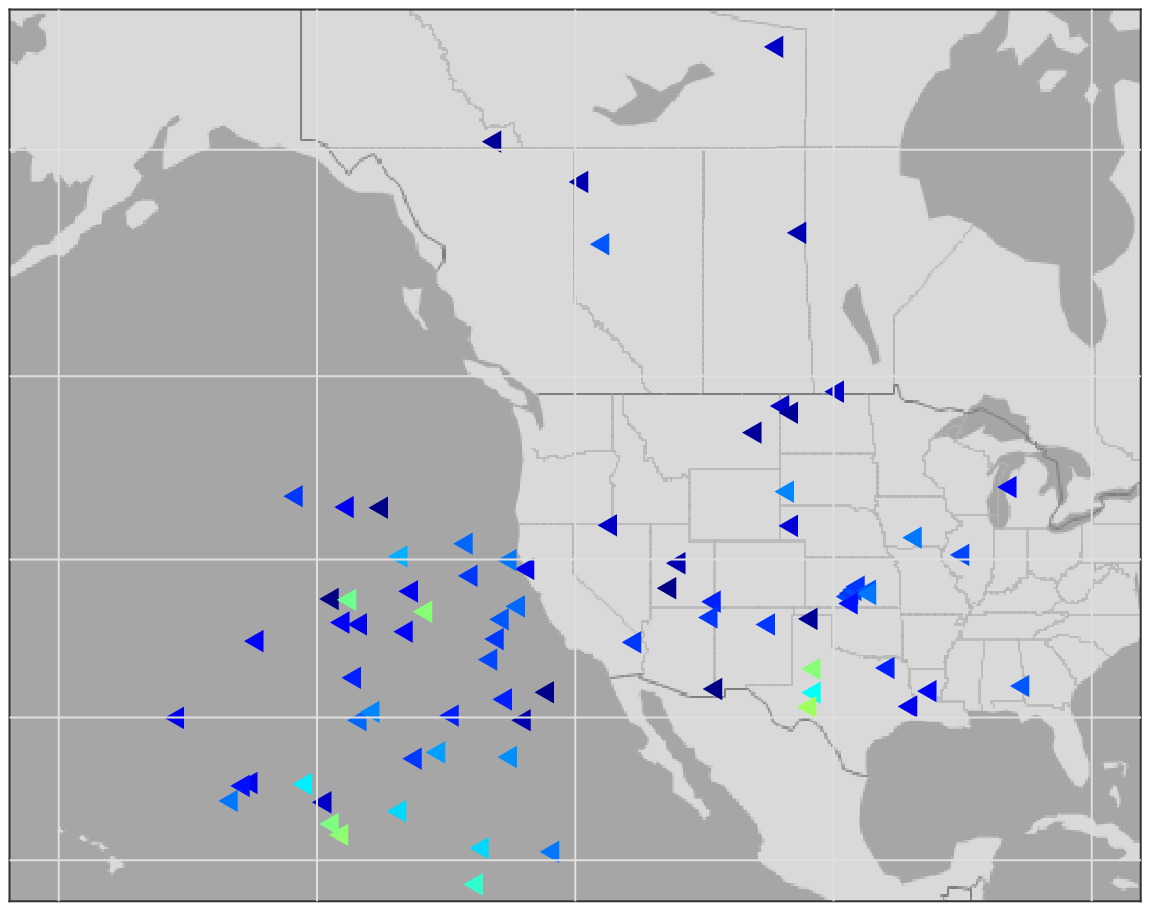}
\caption{41000 ft}
\end{subfigure}
\begin{subfigure}{0.32\textwidth}
\includegraphics[width=1\linewidth]{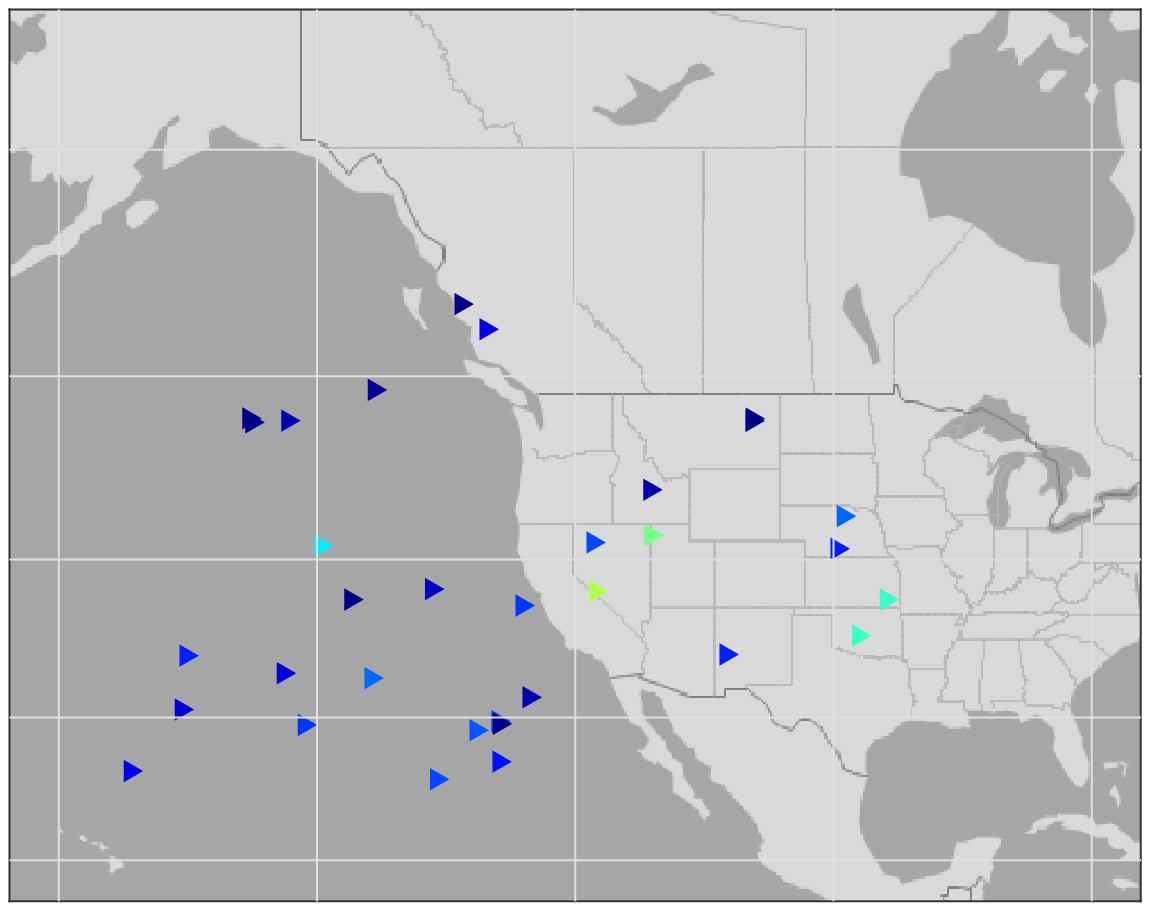}
\caption{42000 ft}
\end{subfigure}
\begin{subfigure}{0.32\textwidth}
\includegraphics[width=1\linewidth]{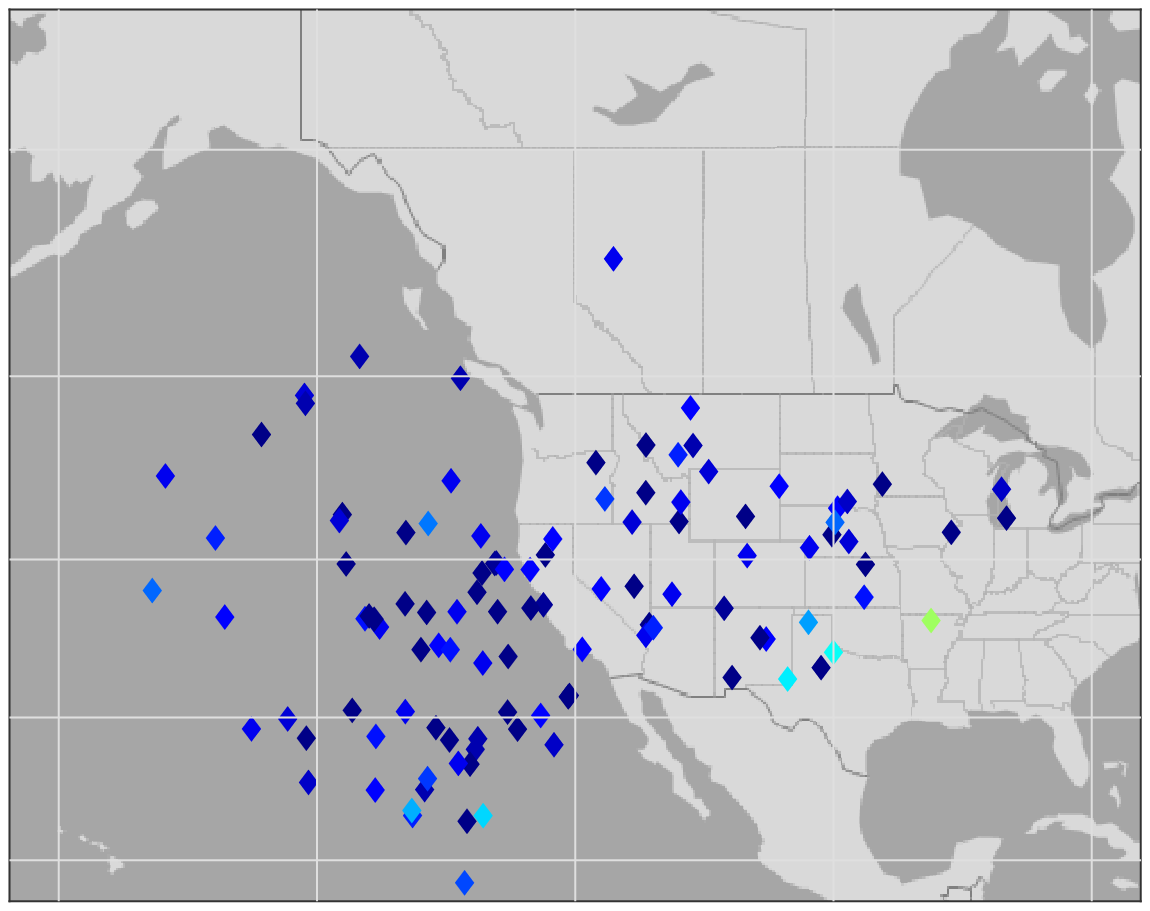}
\caption{43000 ft}
\end{subfigure}

\begin{indented}
\item[]\begin{subfigure}{0.42\textwidth}
\includegraphics[width=1\linewidth]{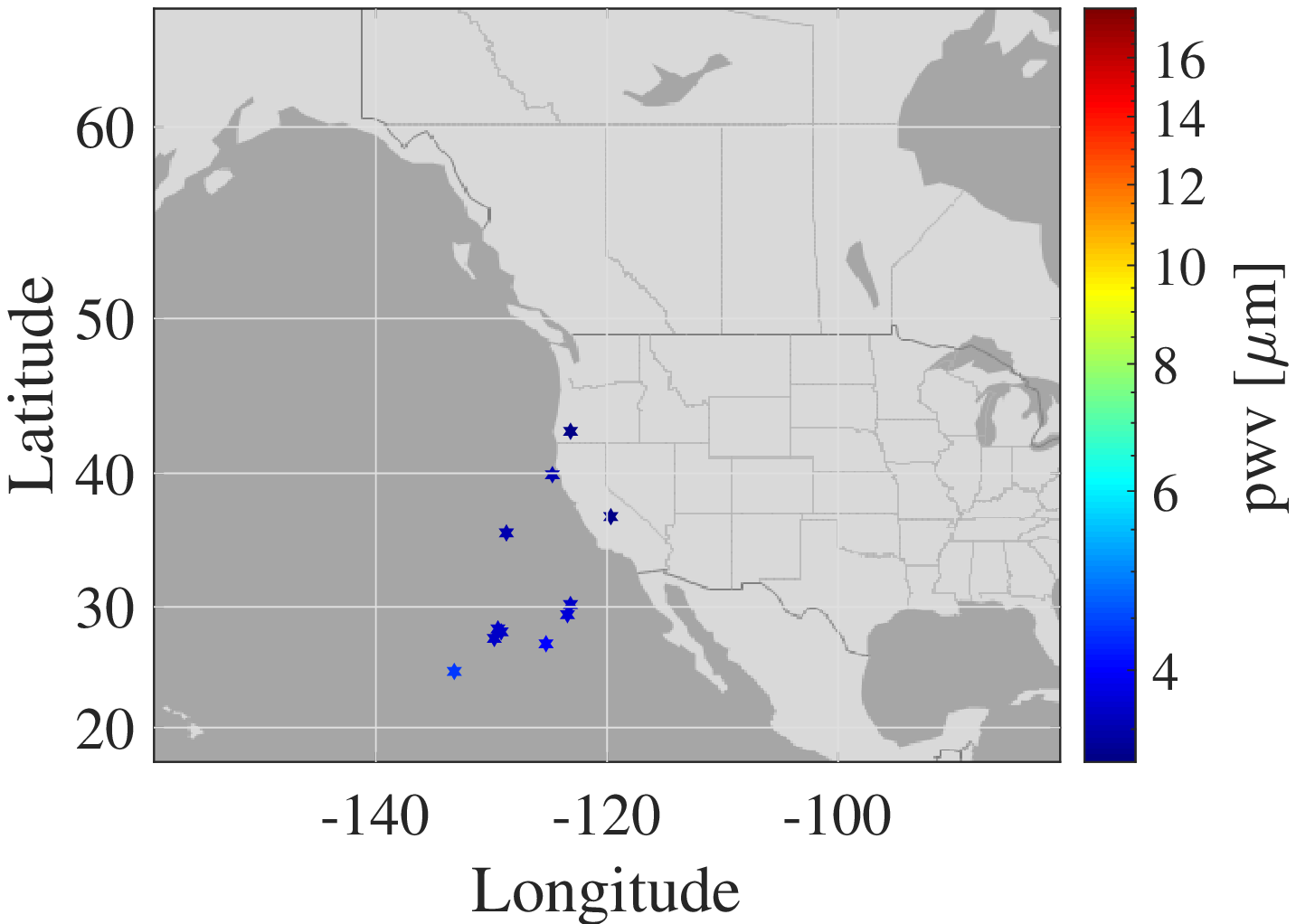}
\caption{$\geq$ 44000 ft}
\end{subfigure}
\end{indented}

\caption{Measured values of water vapor at different pressure altitudes plotted at the position of the measurement.}
\label{fig:maps_altitude}
\end{figure}

Measurements for the different altitudes are not distributed evenly over the area since lower altitudes are typical for the positions closer to Palmdale early in flight, while very high altitudes ($\geq$ 44000 ft) are also also found closer to Palmdale since they are only possible at the end of the flight close to landing in Palmdale. This is shown in the spatial maps in figure \ref{fig:maps_altitude} where the measured water vapor values are shown on the maps separately for the various altitudes. Data points far from Palmdale are typically at 40000 ft to 42000 ft which is the typical altitude halfway into the flight. Positions south of the USA are only possible over the Pacific ocean. This does introduce some positional bias into the sample. Also the plots give an indication how much time SOFIA typically spends at each flight level, with the exception of 38000 ft, where many data points are taken during the setup of the telescope. Also data points become naturally more sparse for the more extreme values of latitude and longitude relative to Palmdale. There are also some atypical data points. The very southern data points at 38000 ft were taken during a long dead leg at the beginning of the flight to position the aircraft after a late takeoff due to a technical issue. So the quite high number of values due to the long dead leg without an astronomical observation represents only a single flight and is of limited statistical use. As already discussed above, problematic values of PWV above 10 $\mu$m are only observed at 39000 ft or below. In addition a trend for lower values with increased latitude is also visible, most notably at 39000 ft and 40000 ft.

\begin{figure}[h]

\includegraphics[width=1\linewidth]{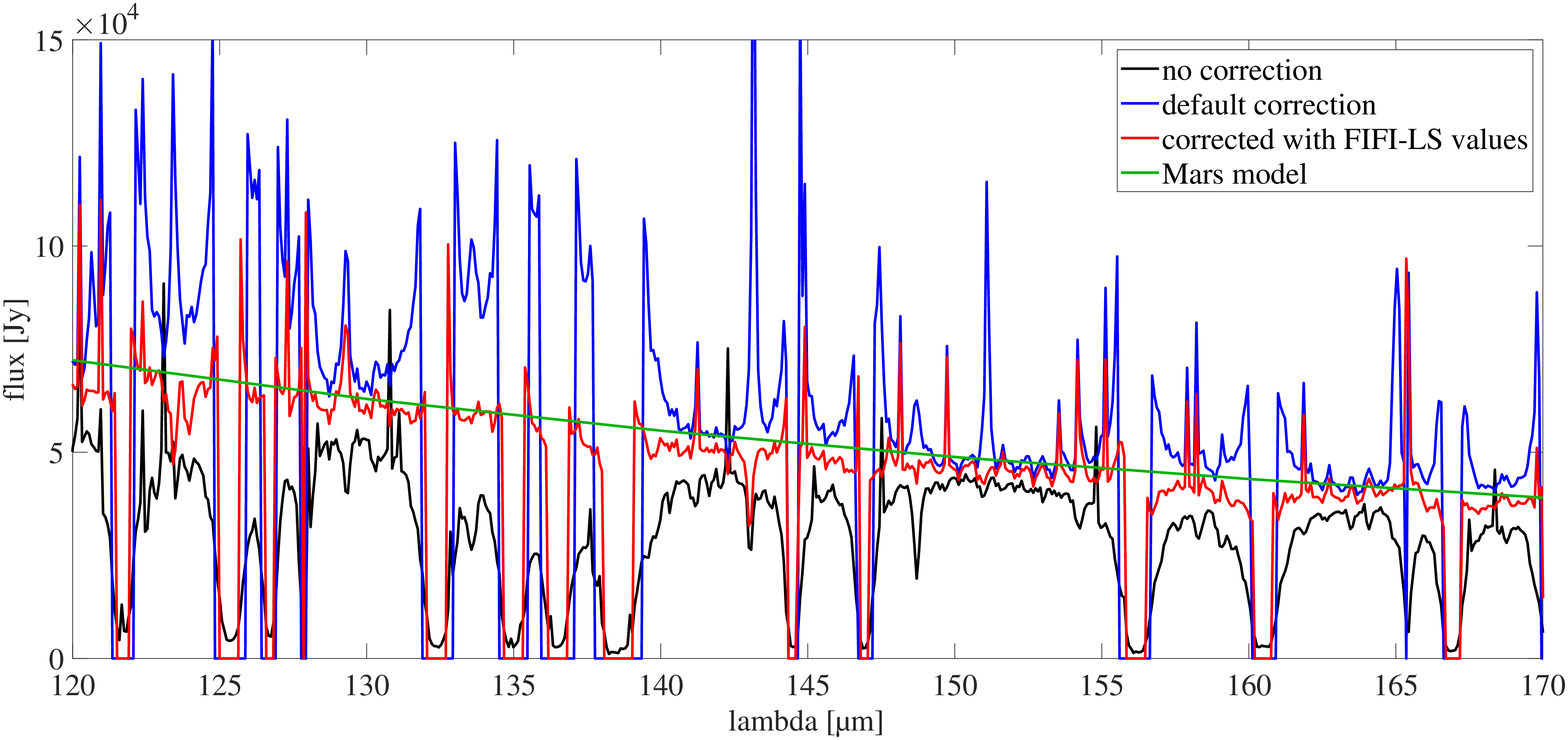}

\caption{Mars spectrum from flight 524 in the red FIFI-LS channel. Shown is the total flux in an aperture with 30 arcsec diameter centered on the spatial pixel with the highest total flux in the cube. The black plot shows the spectrum as measured by the instrument. The blue line is the spectrum corrected with the default PWV value the pipeline uses for the flight altitude (here 11.0 $\mu$m for 38000 ft. For the green line the correction was performed with 6.25µm for the PWV values that was measured by FIFI-LS before and after the Mars measurement.}
\label{fig:Mars}
\end{figure}

\begin{figure}[h]

\begin{subfigure}{0.48\textwidth}
\includegraphics[width=1\linewidth]{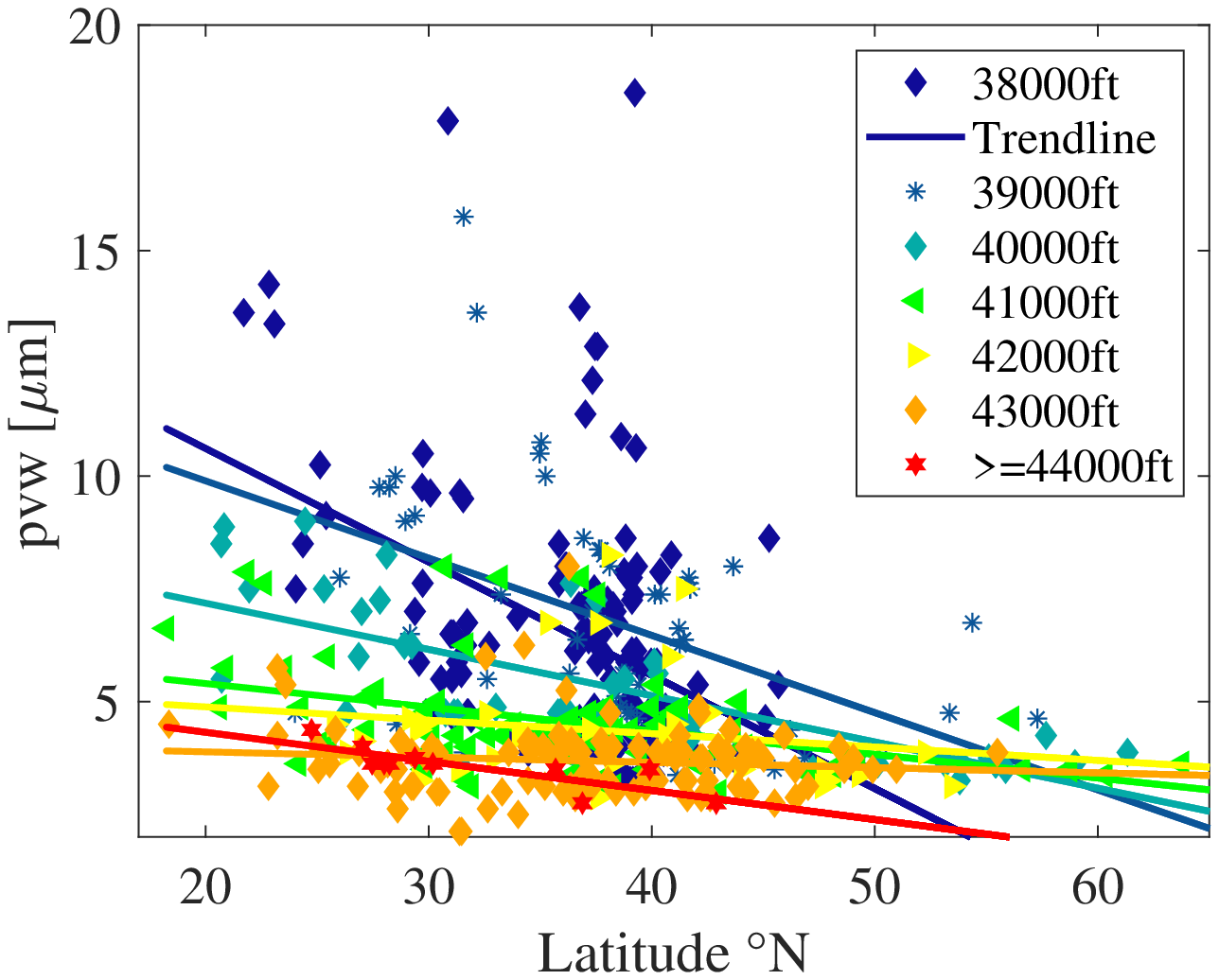}
\caption{}
\end{subfigure}
\begin{subfigure}{0.48\textwidth}
\includegraphics[width=1\linewidth]{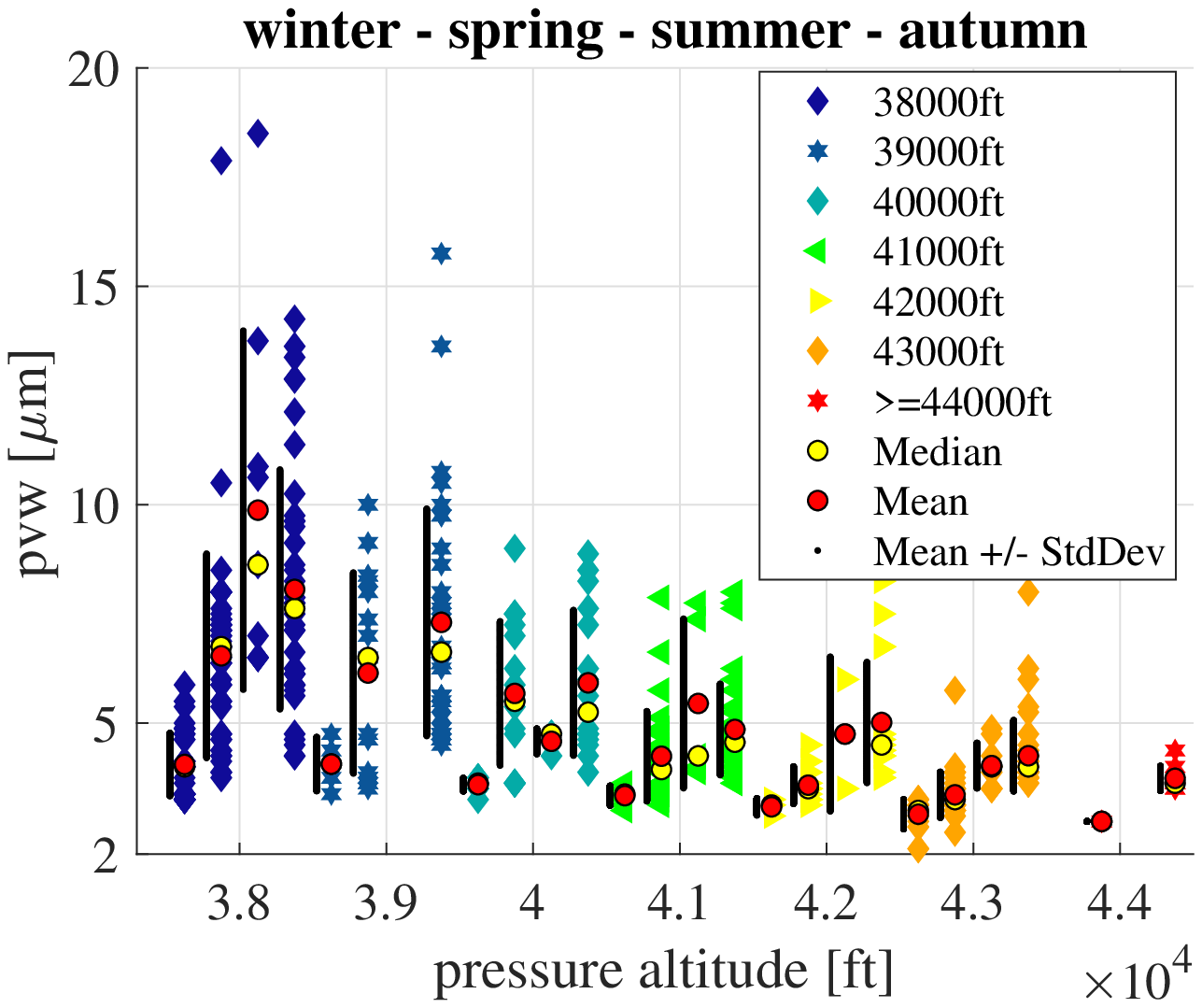}
\caption{}
\end{subfigure}

\caption{Trend analysis of the water vapor measurements: (a) All measured values plotted against the latitude of the position where the data was taken. Different colors indicate the different pressure altitudes. A linear trend line is fitted for each altitude.  (b) All measured values plotted separated into the 4 meteorological seasons as well as  altitudes. Winter is January - March, Spring is April to May, Summer is June to August and Autumn is September to November. For each season/altitude the mean (red point) as well as the median (yellow point) and the standard deviation around the mean (black bar left of the data points) are also shown. There are no data points available for: 39000 ft/summer, $\geq$ 44000 ft/winter and summer.}

\label{fig:trends}
\end{figure}

To show the benefit of measuring the PWV value we use a spectrum of Mars that was taken for the purpose of flux calibration. It is shown in figure \ref{fig:Mars}. Transmission curves for best and worst case PWV values in this spectral range are shown in figure \ref{fig:atm}. The spectrum was taken on flight 524 in a $\sim$35 min spectral scan that contains 86 single integrations to achieve the spectral coverage. The integration was started on the short wavelength side of the spectrum and shifted towards the long end with every integration. The water vapor was measured at 6.25 $\mu$m before as well as after the integration. The data were taken at 38000 ft for which the default PWV value would be 11.0 $\mu$m\footnote{\url{https://dcs.arc.nasa.gov/proposalDevelopment/SITE/index.jsp}}. Figure \ref{fig:Mars} shows the results with the Mars spectrum without any telluric correction in black and a modeled spectrum of Mars in green \footnote{\url{https://lesia.obspm.fr/perso/emmanuel-lellouch/mars/}}. Without any knowledge of the local conditions the data were reduced by the pipeline as shown by the blue curve assuming a water vapor of 11.0 $\mu$m. The pipeline cuts of the signal if the transmission (depending on PWV, elevation and altitude) drops below 20$\%$. Large parts of the spectrum are over-corrected, due to the overestimation of the PWV value. The spikes in the blue spectrum are most prominent around the the spectral regions where the signal is cut off due to low transmission. This causes an overestimate of the source flux up to $\sim$100$\%$. If the FIFI-LS measured value of 6.25 $\mu$m is used, the signal (red curve) agrees with the model (green curve) quite well across of the spectrum, with only a couple of spikes close to telluric absorption features. With the water vapor now obtained by FIFI-LS, the data can be corrected for the atmosphere down to a transmission level of $\sim$20$\%$. This demonstrates that our method works very well for the purpose of correcting SOFIA science data.

In this case the water vapor was measured at the same value before and after the data acquisition and the successful correction of the full spectrum with that value suggests there were no changes of the PWV value during data acquisition. This is not always the case and determining the PWV scale on the scale of minutes can be necessary as demonstrated by \citeasnoun{Iserlohe21:inpress-b} for a similar spectral scan of Mars.

\section{Trend analysis} \label{sec:analysis}

In figure \ref{fig:trends} (a) the water vapor values are plotted against the latitude of the position where the data point was recorded. Since \citeasnoun{1Haas98} demonstrated a clear trend of decreasing PWV values with increasing latitude for all seasons we did not separate those here; combining those cuts together with the necessary separation by altitude yields too few data points for a trend analysis. At each  altitude there is a clear trend of decreasing water vapor with increasing latitude. The trend is strong for 38000 ft and 39000 ft, but the spread of values there is much larger than the trend, so this is dominated by day to day weather variations. Nevertheless, the plot demonstrates that there is an option of optimizing observations by avoiding flight plans that go south at the beginning of a flight. Palmdale is located at $\sim$35$^{\circ}$. All high water vapor values at 39000 ft where observed at $\sim$35$^{\circ}$ or further south. The trend is weaker but still present at 40000 ft, showing the potential gain of avoiding flights that observe from the south at 40000 ft or below. This can be done either by shortening flights to reach higher altitudes early due to the lower weight of the aircraft, or by observing from the north in the first part of the flight before heading south. At the higher altitudes the trend is still present but the overall values are so low that good conditions are almost always guaranteed. Only two  data points at 41000 ft and 43000 ft are available south of 20$^{\circ}$ north and while they are in line with the general trend of increasing water vapor further south, the typical SOFIA flight paths are too far north to check for the significant increase in water vapor overburden predicted in \citeasnoun{1Haas98}.

To check for trends with seasons the measurements for the different altitudes are split up into the meteorological seasons in figure \ref{fig:trends}. For each altitude/season combination the mean and median values as well as the standard deviation around the mean are also shown. From this it becomes clear that the winter season (December to February) has low water vapor values for all altitudes with the highest observed value at 6 $\mu$m, mean and median values at or below 4 $\mu$m and a very small spread of values. As mentioned above data from only 6 flights out of 39 were taken during winter, so the statistical results are not as robust as they are for spring and fall. For the other 3 seasons higher water vapor values do occur. Spring (March to April) has similarly low values compared to Winter at 41000 ft or above but there is an increase in both mean and median PWV values. Also the spread of the values increases, so that observations with $\geq$7 $\mu$m of water vapor overburden do occur. 
Again data points from the summer months are very limited and all were obtained during the second half of August on only 3 flights. In general, both mean/median values and the spread of the data are close to the autumn values but that is likely caused by the data taken only a couple of days before the beginning of September. 
Autumn values are comparable to the spring but consistently higher. The mean and median values are about 1 $\mu$m higher compared to spring. Also the median is lower than the mean at all altitudes indicating a somewhat higher risk of non-ideal water vapor values. Spring has only 2 PWV value above 10$\mu$m at 38000 ft while there are 7 in the fall as well as 5 at 39000 ft compared to none in the spring. Also there are some data points in the fall at 42000 ft or higher with more than 7$\mu$m PWV while there are none present in the spring. The fall sample is larger with 17 flights compared to 12 in the spring. The offset between the mean and median indicates a higher risk of higher water vapor values in the fall compared to the spring. This is consistent with the analysis of \citeasnoun{1Haas98}.

\section{Summary and Conclusions} \label{sec:summary}

We presented water vapor measurements with the FIFI-LS Instrument on SOFIA. Flights were performed in all the seasons out of Palmdale. In general, conditions are favorable, with only a very limited number of data points above 10 $\mu$m. PWV above 10µm was encountered only at 39000 ft or below. Still within expected condition (mean value plus standard deviation) values above 10 $\mu$m at 39000 ft are unlikely both in general as well as when seasons are analyzed separately. 
Conditions appear to be very good in winter at all altitudes and spring shows better (lower) values than fall. But no conditions are expected that would make a SOFIA flight not advisable. The data base in the summer is relatively small making it harder to draw firm conclusions, since SOFIA usually deploys the southern hemisphere during summer. 

We have shown that the PWV value determined by FIFI-LS can improve the quality of the SOFIA/FIFI-LS science data significantly, especially in spectral regions with potentially low atmospheric transmission.  Since our measurements cannot be taken in parallel with science observations the frequency of data acquisition is limited. In this paper we did not discuss the spatial or temporal variations in water vapor during a flight or even a single observation at a fixed altitude. This will be done in \citeasnoun{Iserlohe21:inpress-b}.

We observed a clear trend of decreasing water vapor with increasing latitude at all altitudes. All PWV values above 10 $\mu$m were observed south of 35$^{\circ}$ at 39000 ft or south of 40$^{\circ}$ at 38000 ft. Beginning flights with a heading towards the north would be a possible way to avoid those. Also the general trends to lower PWV values further north does favor northern flight paths for better science data. 

For observations that depend on low water vapor values, there are no limitations in the winter, while in spring and more so in the summer and the fall, scheduling them later in the flight to be observed at 41000 ft or above to keep the expected water vapor below 7 $\mu$m would be helpful. This can be achieved by a shorter flight, but choosing a flight plan farther north can also be helpful to minimize the risk of higher water vapor. 

\section{Acknowledgments}
SOFIA, the "Stratospheric Observatory for Infrared Astronomy" is a joint project of
the Deutsches Zentrum für Luft- und Raumfahrt e.V. (DLR, German Aerospace Centre;
grants 50OK0901, 50OK1301 and 50OK1701) and the National Aeronautics and Space
Administration (NASA). It is funded on behalf of DLR by the Federal Ministry for
Economic Affairs and Energy based on legislation by the German Parliament and funded
by the state of Baden-Württemberg and the Universität Stuttgart. Scientific operation for Germany is coordinated by the German SOFIA Institute (DSI) of the Universität
Stuttgart, in the USA by the Universities Space Research Association (USRA).

\section{References}

\bibliography{harvard}{}

\end{document}